# Machine Learning and Semantic Analysis of In-game Chat for Cyber Bullying

Dr Shane Murnion, Prof William J Buchanan, Adrian Smales, and Dr Gordon Russell[1]

**Abstract—** One major problem with cyberbullying research is the lack of data, since researchers are traditionally forced to rely on survey data where victims and perpetrators self-report their impressions. In this paper, an automatic data collection system is presented that continuously collects in-game chat data from one of the most popular online multi-player games: World of Tanks. The data was collected and combined with other information about the players from available online data services. It presents a scoring scheme to enable identification of cyberbullying based on current research. Classification of the collected data was carried out using simple feature detection with SQL database queries and compared to classification from AI-based sentiment text analysis services that have recently become available and further against manually classified data using a custom-built classificationclient built for this paper. The simple SQL classification proved to be quite useful at identifying some features of toxic chat such as the use of bad language or racist sentiments, however the classification by the more sophisticated online sentiment analysis services proved to be disappointing. The results were then examined for insights into cyberbullying within this game and it was shown that it should be possible to reduce cyberbullying within the World of Tanks game by a significant factor by simply freezing the player's ability to communicate through the in-game chat function for a short period after the player is killed within a match. It was also shown that very new players are much less likely to engage in cyberbullying, suggesting that it may be a learned behaviour from other players.

## I. INTRODUCTION

Cyberbullying is an anti-social behaviour that can be defined as repeatedly and intentionally causing harm to others using computers, cell phones, and other electronic devices and has been shown to have negative outcomes for its victims including depression, stress, poor performance in school and work and in extreme cases suicide. One area affected by cyberbullying is online gaming which has grown rapidly over the last few decades, achieving worldwide popularity with hundreds of millions of users, half of whom report that they have been victims of cyberbullying at some point in time. cyberbullying causes even financial concerns since these toxic activities represent a serious cost to the providers of online gaming services that are part of an industry sector worth more than 100 billion dollars in 2016.

### A. *On-line multi-player gaming and World of Tanks*

The social activities of young adults increasingly take place on the Internet: within social media, through digital communities such as Reddit and within online multi-player games (Homer et al., 2012; Subrahmanyam and Smahel, 2011). Online gaming has grown an enormous player base as Table 1 shows, with popular games having many millions of users per month (NowLoading, 2016; PCGameN, 2016; Wargaming.net, 2017a).

Beyond these traditional forms of online gaming there had also been a huge growth in games built-in to social media (Entertainment Close-up, 2012; Business Wire 2015). Social interaction with other players has become an important aspect of many online games, whether it is cooperative play in Massively Multiplayer Online Games/Online Roleplaying Games (MMOG/MMPORG) such as World of Warcraft or World of Tanks, or just social play within traditional social media such as Facebook. Gaming has traditionally been more popular among men but research indicates that gender differences in this area are diminishing (Homer et al., 2012) so that gaming has become a major leisure time activity for both boys and girls.

| Game | Worldwide Popularity | User base |
|---|---|---|
| League of Legends | 1st | Over 100 million per month |
| Hearthstone | 2nd | Over 50 million per month |
| DOTA 2 | 4th | Over 13 million per month |
| World of Tanks | 8th | Over 1 million users at any point in time. |

**Table 1 Userbase for popular online multiplayer games**

World of Tanks (WOT) is a massively multiplayer online game (MMOG) where 2 teams consisting of up to 15 armoured vehicles compete on a battlefield with the goal of capturing a base or annihilating the enemy team. The players in each team can either be randomly chosen or consist of players who have chosen to play together temporarily (a platoon) or as part of a permanent group (a clan). During the match, all the players can communicate with other players on their team only via in-game chat messages until the match is complete. When a player is killed they can choose to leave the match or alternatively to stay; observing the game and chatting with other (both living and dead) players on their team. Previously in-game chat with the opposing team was also possible, but this feature was removed from the game in 2016 specifically to help reduce toxic behaviour and abusive language (Wargaming.net., 2017e). The game has a number of persistent elements; players have an in-game identity that accumulates experience points which can be used to unlock new and more powerful vehicles and gain new abilities for the tank crews and indeed years of play may be required to unlock the top-level vehicles in the game. Another persistent element is the publicly available player statistics such as percentage of matches won. Improving these statistics is the main goal of many players in the game. A final important persistent element is game replays which are made available through third party services allowing players to publish matches that can be viewed by the game community at large.

[1] Dr Shane Murnion and Adrian Smales, are researchers in The Cyber Academy, Edinburgh Napier University. Prof William J Buchanan and Dr Gordon Russell are academics The Cyber Academy, Edinburgh Napier University.

### *B. Contribution*

This paper defines a method to automatically collect in-game chat data from WOT matches in a sustainable manner through web scraping of a match replay website combined with Extract, Transform and Load (ETL) techniques to build a new database of in-game chat messages and to demonstrate the potential for data gathering. The collection system also identifies players and fetches player information from the public API services published by Wargaming.net, the publisher of World of Tanks, to provide extra value around the chat data collected.

Having collected the data a prototype classification client was built and tested to enable the rapid classification of chat messages, in an attempt to evaluate the possibility of creating training data useful in machine learning analyses. Further a simple analytic exercise was carried out to demonstrate the utility of the data collected. A technique based on artificial intelligence known as sentiment analysis was then examined as a possible tool for automatic detection of cyber-bullying chat messages.

### *C. Motivation*

The paper has a number of goals within its scope:

- **Provide data that is free.** This was successful, the data will be made publicly available to researchers, though a particular source has not been chosen at this time, possibilities include Kaggle or a custom solution using an Azure SQL database.
- **Dynamically collect data continuously over time, rather than generate a snapshot of one point in time.** This goal was also reached. The data collection system runs automatically as a scheduled windows service and the figures shown in Table 3. The goal is to create a dataset with hundreds of millions of chat messages over many years.
- **Allow the identification of players over time and between games.** This is already possible, since gamer identities in World of Tanks rarely change. One interesting avenue of research here would be to follow individuals who engage in cyberbullying activities and study how that behaviour changes as they progress in ability and even simply get older and become more (or less?) mature.
- **Guarantee the anonymity of the players within the dataset.** This is not the case today, however the primary key in the WotPlayer table is a GUID and not the player's username. Before publication even these player identities will be removed from the publicly available data so that the database is fully anonymous in every sense. The data collection system will however maintain a private database with mappings between the user's game identity and their GUID. This will be carried out to enable longitudinal studies on particular players. It should be noted that the GDPR does not protect game ids but there is the small possibility that somehow a gamer identity could be linked to a real-life identity and then the chat data within the game could be used to negatively affect that person. The possibility of this happening will be removed before publication of the data.
- **Provide useful attribute beyond just chat data where possible.** This was successful but only to a very limited degree, in that information about how long the player has played, how much experience they have and their level of ability was gathered. Other information that might be useful to researchers, such as age, sex, nationality, education and other soft factors were not available. These would however be very interesting if they were accessible. Wargaming.net should have some of this information in their internal systems, however it is unlikely that they would release this information even for research purposes.
- **Contain data spread geographically over the globe.** The system currently logs data from both the American and European servers. It would be interesting to log the information from Russian and Asian servers as well, giving global coverage and allowing many sorts of interesting geographical analyses. However, in order for this to be successful some sort of translation system would have to be used to covert from Russian, Thai, Korean etc. to English. This is very feasible today with online services, the only real obstacle is the price tag. Prices for these types of services are falling and so it may be feasible within the near future.
- **Contain data covering different game systems.** This was outside of the scope of the paper, but would of course be an important next stage in the research. Since the game companies themselves have an interest in combatting in-game anti-social behaviour it might be possible to come to some working arrangement with the companies to gain access to their data for research purposes where the results may well be helpful to them in turn.

## II. Literature Review

### *A. Cyberbullying: the dark side of social gaming*

It is perhaps not so surprising that in parallel with the growth of social activity within gaming there has also been a growth in anti-social activities behaviours (Kwak et al., 2015). Types of anti-social or disruptive behaviour (often referred to as "toxic" within the gaming community) include "griefing", chat spamming, bug exploitation, and cyberbullying (including racial or minority harassment). Although these concepts are separate there is a degree of overlap in the definition of some of these behaviours i.e. a griefer has been described as a player who derives enjoyment within a game by reducing the enjoyment of other players (Mulligan et al., 2003), while chat spamming is a disruptive technique that floods the in-game chat with some text, often the same text, repeated over and over, which effectively blocks the communication channel and distracts players (Hinduja and Patchin, 2008). Thus, while griefing and chat spamming are separate and somewhat different, nonetheless, a griefer could employ chat spamming as one technique in his or her arsenal.

A dominant single definition of cyberbullying has yet to emerge although many contain similar or common elements. One definition of cyberbullying focuses simply on the behavior e.g. being cruel to others by engaging in socially aggressive behaviour using the Internet or other digital technologies (Hinduja and Patchin, 2008) and since this work is also focused on the same theme this is the definition used here. It should be noted however that other definitions place stricter requirements before the activity can be considered cyberbullying as opposed to the broader definition of online aggression: the repetition of the behavior over time, an imbalance of power between the perpetrator and the victim and finally the intention to do harm (Giménez Gualdo et al., 2015; Smith et al., 2008), although the importance of some or all of these aspects is questioned (Cuadrado-Gordillo, 2012; Slonje et al., 2013a). A further active topic of debate is how to define repetition or even power imbalances in online activity as opposed to normal bullying (Slonje et al., 2013b). There are many examples of the grey areas here: is a retweeted message multiple malicious acts or one? Is a humiliating video uploaded once but viewed many times by the victim repeated bullying? Who has more power in gaming activities, is it the better player with a better reputation, the wittiest player with the best put-downs, or as some suggest (Vandebosch and Van Cleemput, 2008) the user with the best grasp of technology The competitive nature of many modern games (as demonstrated by the current growth in E-sports), the inclusion of in-game chat features and use of persistent in-game identities are all factors that combine to create a fertile environment for the occurrence of cyberbullying (Kwak et al., 2015). The problem of cyberbullying has become so prevalent that it has featured on the UK national news:in 2017 the BBC reported on a survey carried out by Ditch The Label, an international anti-bullying charity, that 50% of young gamers had experienced cyberbullying (BBCNews, 2017; Ditch The Label, 2017).

The consequences of cyberbullying in general can result in number of negative outcomes such as poor performance in school (Hinduja and Patchin 2008), fear (Giménez Gualdo et al., 2015), stress, loneliness and depression (Ortega et al., 2012). It is not just a problem for children either, cyberbullying has shown to have negative effects in the workplace too (Daniels and Bradley, 2011; Moore et al., 2012). In extreme cases cyberbullying has been linked to trauma, suicide and acts of violence (Daniels and Bradley, 2011; Moore et al. 2012). Specifically within gaming the negative effects of cyberbullying can be seen for the game vendors for whom the financial consequences of cyberbullying can be severe (Mulligan et al., 2003; Pham, 2002), indeed 22% of players in the Ditch The Label survey reported that they had stopped playing a game as a consequence of cyberbullying, while a further 48% had considered quitting (BBC News, 2017). Another study reports similar results: 38% of players had avoided a multi-player game over concerns about cyberbullying, 54% had left a match because someone was exhibiting this type of behavior and over 63% agreed or strongly agreed that cyberbullying is a serious problem in the online gaming environment (Fryling et al., 2015). This is a major consideration for the gaming industry which was worth over one hundred billion dollars in 2016 (McDonald, 2017) to whom it is clearly important that solutions are found (Balci and Salah, 2015). Consequences exist for gamers on the personal level as well. Studies have shown links between cyberbullying activity and aggressive behavior in real life (Yang, 2012) and even increased isolation and dimished self-esteem (Fryling et al., 2015). Indeed there have been several cases where the negative behavior within a game has resulted in revenge actions within real life violence and threats (Mail Online, 2015, 2011).

The industry is aware of the problem and has attempted various solutions. League of Legends, one of the most popular games in the world today instituted a system where players with anti-social chat behaviour could be judged as such by their peers in a system called “The Tribunal” with subsequent punishment if the player was found guilty (Kwak and Blackburn, 2014). However Riot Games Inc. the owners of LOL apparently were not satisfied with the solution since the system was taken down several years ago “for maintenance” and has yet to reappear (ElektedKing, 2015). Other game systems have applied simpler solutions such as allowing players to block all communication or filter out swear words.

The problem with blocking all communication is that in-game communication can be very useful to a player, allowing for requests for help, or planning an attack or a defence. In the case of World of Tanks abusive chat messages often come from players who have died and are upset. These dead players remain in the match as observers but unfortunately use the same chat channel as the living teammates that a player still needs to coordinate with.

Filtering text is also of limited value since there is not much difference between the filtered phrase “Just die you useless piece of ****” and the unfiltered version. Both are likely to cause a certain amount of stress to the recipient.

A more sophisticated nuanced solution is needed that can recognise and remove bullying text (or take some other action such as warning or reporting the perpetrator) while allowing through more normal and useful communications.

### B. *Cyberbullying and the law*

In the UK there is as yet no specific legislation dealing with cyberbullying, however the actions that are involved in this type of behaviour are often covered under current laws. Section 1 of the Protection from Harassment Act 1997 prohibits persons from carrying out acts that he or she knows amounts to harassment in one way or another (UK Gov, 2017). Section 2 of the act is a less serious version of the offence but is still punishable with up to six months in prison or up to £5000 in fines. A more serious offence is defined as repeated occurrences of the offence (on at least two occasions) and entails a fear of violence. This offence can lead to up to five years in prison, fines or both.

The importance of repetition in the law here is noteworthy, matching as it does the requirements of some definitions of cyberbullying. Other researchers have pointed out that several other laws may be applicable in cases of cyberbullying including the Communications Act 2003; the Malicious Communications Act 1988; the Telecommunication Act 1984; the Public Order Act 1986; the Obscene Publications Act 1959; the Computer Misuse Act 1990; the Crime and Disorder Act 1998 and the Defamation Act 2013 (El Asam and Samara, 2016). In World of Tanks, for example, where it is not unknown for players to receive death threats against them and their families (ArmouredBin, 2015; Hjalfnar_Feuerwolf, 2017; Militiae_Cristianae, 2015; Psychovadge, 2014), such actions may contravene either the Offences against the Person Act 1861, or where the intent in the threat is doubted, section 4 of the Public Order Act 1986 (Crown Proecution Service, 2017).

Internationally the law is changing in this area and cyberbullying (as well as traditional bullying) is specifically illegal in over 20 states of the US such as, for example, Louisiana (Patchin and Hinduja, 2015). Given developments in this area it seems likely that cyberbullying will be shortly become a defined cybercrime within the UK if it is not already so implicitly.

### *C. Research into cyberbullying within online games*

Considering the serious potential consequences of cyberbullying, the huge growth in online gaming, the growth of social gaming within social media, the development of social activity within gaming and the concomitant development in anti-social activities one might expect that research into cyberbullying within the gaming world would be an active area. In reality despite the intense research in Facebook, social media and online communications in general, the gaming aspect is often neglected by researchers in the fields of communication research in general (Wohn and Lee, 2013). Research within cyberbullying has tended to focus on messaging (Moore et al. 2012), email, chat rooms (Smith et al. 2008), online forums (Moore et al. 2012) and traditional social media such as Facebook (Kwan and Skoric, 2013) or MySpace while research within gaming has tended to focus on griefing rather than cyberbullying e.g. (Foo and Koivisto, 2004) have examined the intentions of griefers, while (Chen et al., 2009) have investigated if anonymity and immersion are contributing factors to griefing. The focus on griefing may be a consequence of the very real financial cost to online games generated by griefing behaviour (Mulligan et al., 2003; Pham, 2002).

Although it is often overlooked, some research focusing directly on cyberbullying within gaming can be found. One study has examined whether or not gender is an issue within in-game cyberbullying and if perpetrators of bullying tend to have been victims of cyberbullying themselves (Yang, 2012).

Cyberbullying researchers have traditionally relied on surveys and questionnaires exploring the victim's or perpetrator's own experiences(Bauman et al., 2013; Giménez Gualdo et al., 2015; Kwan and Skoric, 2013; Lam and Li, 2013; Pettalia et al., 2013; Smith et al., 2008). There is very little in the way of concrete data that contains the actual cyberbullying text and where such data does exist the focus has tended to be within traditional social media (Reynolds et al., 2011). Taking an alternative approach from such traditional studied (Kwak and Blackburn, 2014) carried out a large-scale analysis of cyberbullying and other toxic behaviours within a gaming context using freely available online data generated by the gaming company itself. This gave the researchers access to an extremely large dataset with 590,000 cases from the League of Legends game available through the "Tribunal" system (TenTonHammer, 2017) where chat data for heavily reported players was made available for judgement to the player community and it was this data that the researchers used to linguistically analyse in-game data (Kwak & Blackburn, 2014).

The lexical analysis and timeline analysis was carried out using the LOL dataset to see if one can identify text that is associated with cyberbullying or even if one can examine temporal patterns of in-game chat to determine when a player's behaviour changes from normal in-game chat to toxic cyberbullying activity. The study also attempted to extract a lexicon of words that could be associated with cyberbullying and used as part of some automatic detection or warning system and suggested that this is a key area for future study. There is an inherent problem with such linguistic prediction; in that it is difficult for any system to differentiate between innocuous *trash talk* which may use the same words as cyberbullying toxic chat.

The dataset used did have some drawbacks. The data was anonymized so that it is impossible to identify serial cyberbullying where it occurs over several games, and furthermore was biased towards toxic behaviour in that only chat data from potential cyberbullies was made available. Thus, it is hard to compare a "normal" match with a match where toxic behaviour had potentially occurred since only potentially toxic match data was made available. Unfortunately, the tribunal system has been closed since 2014 (ElektedKing, 2015) and is still currently unavailable. Nonetheless is unfortunate that such a wonderful resource for cyberbullying researchers is no longer being renewed. The dataset collected previously is however still being used and recently the researchers have by examining the same team-based game (League of Legends) examined the possible origins of in-game cyberbullying: whether or not anonymity and team performance played a role and if conflicts tended to be intra-team or between teams (Kwak et al., 2015).

This more recent research has started to define some important goals for anti-cyberbullying research: is it possible to build systems that can detect, counter and perhaps even prevent cyberbullying? Looking at the first of these goals, the detection of cyberbullying it is clear that in order to implement cyberbullying detection there must exist a quantitative definition of what cyberbullying is, some sort of measure which by which we can define the likelihood that a behaviour is in fact toxic behaviour or not, a cyberbullying intensity score.

### *D. Measuring cyberbullying intensity*

Studies of cyberbullying and normal bullying often in fact attempt to quantify the extent or scale of the bullying by attributing some sort of score. Most scales of measurement are based on the frequency or the duration of the bullying activity (Bauman et al., 2013; Giménez Gualdo et al., 2015; Yang, 2012). An alternative way to score the bullying activity has been by the reaction or the perceived reactions of the victims themselves (Pettalia et al., 2013).

Other studies have attempted to examine cyberbullying in more detail looking at the frequency of different types of cyberbullying activities, examining the anti-social activities in more detail e.g. “posted mean comments” or “threatened someone online” (Hinduja & Patchin, 2013; Lam & Li, 2013; Rivers & Noret, 2010). This type of behaviour separation as a tool for analysis has also been used with tradition bullying, for example, Roos et al. (2011) quantified bullying by the physical or verbal activity involved: identifying six different behaviours: *uses physical force to dominate*, *gets others to gang up on a peer*, *threatens others*, *when teased*, *strikes back*, *blames others in fights*, and *overreacts angrily to accidents*. Bullies were then given an overall score based on the number of bullying behaviours they engaged in out of the possible six. In internet media where physical actions are not relevant and verbal or written actions dominate, then lexical scoring might be useful. Reynolds et al. (2011) quantified message-based cyberbullying by scoring each post on the number and severity of the swear words within the post.

If lexical content analysis examines the *what* within cyberbullying analysis then perhaps another useful way to examine the behaviour is by asking “when?”. Timelines of activities within cyberbullying can also be of interest. be another. Several researchers emphasise the importance of repetition as possibly the most important aspect of bullying to the extent that some only define the anti-social behaviour as cyberbullying if it is a repeated activity (Giménez Gualdo et al., 2015; Lam & Li, 2013; Patchin & Hinduja, 2015). In this sense, an individual toxic behaviour can be identified in isolation, but is more likely to be important when identified as part of a repeated pattern of behaviour. In traditional bullying, these timelines can extend over days, weeks or years, but for cyberbullying the timeline is more likely to be compressed down to a single match. This does not change the significance of repetition. One player making a negative remark about another player is probably a much less significant cyberbullying event than a player repeatedly hounding and distracting another player throughout an entire game.

### *E. Potential solutions to cyberbullying*

Potential solutions to the cyberbullying problem have been suggested and studies have shown that parental involvement can mitigate the extent of cyberbullying (Hinduja and Patchin 2013) but the ubiquitous access to social media and online communities through computers, game consoles, mobile phones etc. makes it increasingly unlikely that parents can provide the continuous adult supervision to the extent needed. One possibility is the use of automatic threat identification that could alert a responsible adult, warn an offending player or take a more active automatic role e.g. filtering out communication that constitutes cyberbullying or other anti-social behaviours (Moore et. al. 2012).

The problem of identifying intent from textual messaging is not trivial and Machine Learning has been identified as a possible tool for use in cyberbullying by many researchers. In a recent real life situation machine learning techniques were used to identify the real life persons behind online trolling activities within a school (Galán-GarcÍa et al., 2015). Machine learning techniques have also been applied to cyberbullying detection within twitter (Al-garadi et al., 2016), Ask.fm (Raisi and Huang, 2016) and other social networks (Chavan and S, 2015; Reynolds et al., 2011). (Balci and Salah, 2015) have examined automatic aggressive/abusive behaviour detection in online Okey games using machine learning techniques with promising results. As such machine learning must be considered as a candidate for cyberbullying detection. One area of machine learning and AI that has recently surfaced may be of particular interest when the classification of text is considered and that is the new area of sentiment analysis.

Supervised machine learning techniques require good general datasets to build systems capable of properly classifying unseen data. Unfortunately, data relevant to cyberbullying in gaming environments is not always easy to come by.

### *F. Sentiment Analysis*

Sentiment analysis, sometimes referred to as opinion mining or emotion AI, is a branch of artificial intelligence research that attempts to quantify emotional labels, positive or negative attitudes (polarity classification) and emotional intensity from text. This is used to determine the attitude or emotional reaction of a writer (or another object) to some topic, object or event.

Researchers started using Natural Language Processing (NLP) to interpret attitude within text (Qu et al., 2004) and the interest in the area has grown with the importance of social media, where sentiment analysis is seen as an important business intelligence tool for customer relationship and brand management, allowing companies to discover whether or not customers have positive or negative attitudes to their products (Cambria, 2016). But sentiment analysis is capable of much more nuanced analysis than simply a tool for solving polarity detection problem of positive and negative attitudes (Cambria et al., 2017).

Sentiment analysis has been shown to be useful in detecting irony and sarcasm from text (Farias & Rosso, 2017) and opinion target detection (Peleja & Magalhães, 2015). This last is of interest within in-game cyberbullying where a negative message might have more impact when it is directed against an individual: e.g. *player X is useless*, than when it is directed against a group, e.g. *this team is useless*. Semantic analysis techniques have already been considered for use in this general area e.g. for filtering out trolling messages and spam in online communication (Cambria et al., 2017). In this paper, an attempt was made to use AI techniques to automatically classify messages by attitude to see if this could be helpful in automatically detecting cyberbullying.

### *G. The search for data*

One important goal that has arisen from current cyberbullying research: that of building a system that can detect, counter and perhaps even prevent cyberbullying? In order to detect cyberbullying in practice there must exists a quantitative measure of what cyberbullying is.

Machine learning is identified as a possible candidate technology that would be useful in building cyberbullying detection systems and this generates other requirements. Supervised machine learning techniques require a classified dataset that can be used to *train* the algorithm. In order to be generally applicable such data would have to be representative of the type of data the system would be required to classify. In gaming terms, this means that data is needed from several different online games if there is to be any possibility of creating a general classifier. The data sets should be large, to provide diversity of data, and continuously evolving since terms of speech and patterns of social interaction can change quite rapidly where the internet is concerned.

An important question is therefore where will this data come from? In an era of big data researchers should not be limited by surveys and questionnaires, although such tailored data will no doubt continue to be useful. picture is not terribly positive, since many of the datasets used are limited in size and scope and even where large datasets were used problems exist. The large dataset, for example, used in studying behaviour in League of Legends by Kwak & Blackburn (2014) is no longer available and the large dataset used to examine the Okey online game is proprietary (Balci and Salah, 2015).

Some research data with general cyberbullying/machine learning focus are available (Edwards, 2012; SwetaAgrawal, 2017; Wisconsin-Madison, 2016) but it becomes more problemtatical to find data concerning cyberbullying and online gaming. Kaggle has become a popular repository of publicly available datasets that can be used in machine learning (Garcia Martinez and Walton, 2014; Kaggle, 2017). A quick search of the Kaggle repository reveals DOTA 2 game datasets, one at least of which contains in-game chat data from 50,000 DOTA 2 matches (Kaggle, 2017). It is a positive sign that any data is available but one static dataset for one particular game hardly meets the requirements for the goals mentioned here. Much more data will be required over many gaming systems if general automatic solutions are to be found that work with new games as soon as they appear and the scarcity of detailed datasets like these represent a direct hinder to cyberbullying research and the search for solutions. The potential for data collection is enormous since there are several popular online games that are both extremely popular and notorious for the toxicity of their in-game chat including League of Legends (LOL) (‘League of Legends’, 2017), Defence of the Ancients 2 (DOTA2) ('Defence of the Ancients' 2, 2017) and World of Tanks (WOT) ('World of Tanks', 2017). These games generate an enormous amount of in-game chat every day, both toxic and otherwise, and represent an enormous pool of potential data for researchers.

## III. Data Collection System

### *A. Data sources relevant to WOT in-game chat messages*

The system for collection of WOT in-game chat messages used two major data sources of WOT data. The first data source was binary WOT replay files, mostly obtained from WotReplays.com which contain the actual match/game information: the players involved, the match results, time of death of each player and the messages from the in-game. The second data source was the WG public API, a full set of services (WOT API) which provide information and statistics on players within the World of Tanks game e.g. how experienced the player is, how many battles they have engaged in, their success rate, and so on. The combination of these data allows for the evaluation of various hypotheses or research questions e.g. Are players more likely to engage in hostile chat behaviour immediately after dying? Or are beginner players less likely to engage in cyberbullying than experienced players? Even vehicle type was included since certain types of vehicle are less popular than others in World of Tanks and thus a player using a certain type of vehicle may be more likely to be the target of cyberbullying. One good example of this is the artillery class of vehicles which are seen to be quite divisive according to the player community in WOT (Garbad, 2014; Granducci, 2016).

### *B. World of Tanks In-Game Data from Replay Files (WotReplays.com)*

When a person plays a game in World of Tanks, the game data from that match is saved to a replay file (Wargaming.net, 2017b) in a binary encrypted format. These replay files allow the user or other persons who have access to the file to replay the match and have a number of uses e.g. teams will often check replays to improve their game play or YouTube contributors may use replays to share a particularly good game with their audiences. Players can submit a replay file to a particular website “WotReplays” (WotReplays, 2017) to be shared by other players in the community. Thousands of replays are uploaded to WotReplays every day from the four WOT player communities: North America, Europe, Russia and Asia. WOT replay files contain detailed information on each match, the players involved, actions taken by the players and even the in-game chat. The WotReplays website was thus a potentially rich source of WOT in-game chat data and in this paper a system was built to download the replay files from the WotReplays website, extract useful data from the files and store it in a useful format within a database so that it could be made available to the general research community.

#### *Replay File Encryption*

Wot replay files are encrypted using the Blowfish encryption algorithm (Nie and Zhang, 2009), luckily the interest in WOT replay files is such that there are several available tools from GitHub that can decrypt, extract and convert binary WOT replay file data to a more useful form such as JSON data such as WoT Replay Analyzer (Aimdrol, 2017), wotreplay parser (Temmerman, 2017) and wotdecoder (Rasz_pl, 2017).

#### *World of Tanks Player Information from the WG public API*

Wargaming.net the company responsible for World of Tanks encourages the development of third party applications that support game play. To support this development the company has made available a wide range of web services (WOT API) providing data on player statistics (Wargaming.net, 2017c). Since the WOT replay files contain the unique Id of the players involved in that match, this Id information can be used to query the WOT API to obtain game statistics for those players, to give a fuller picture of the players who produce the in-game chat messages embedded within the replay file.

### *C. Data collection process*

The data collection used in this paper was based around a custom-built web crawler/web spider (Wikipedia, 2017a). This spider navigates through the web pages on the WotReplays.com website and “scrapes” the web pages(Wikipedia, 2017b) collecting links to downloadable WOT replay files. The files discovered are downloaded and converted from their encrypted binary format to standard JSON text files using the WOT replay parser tool. The second stage of the process is the parsing of the replay data. Files from servers, other than the European or North American servers were rejected since the focus on this paper was cyberbullying using the English language. Suitable files (from the European or North American servers) were parsed in three separate stages.

*Parsing the data stage 1: general replay information*

The general information and meta data about the replay was extracted e.g. when did the match take place and this metadata is persisted to the data store. The date of the match is important since attempts were then made to obtain the current statistics of the players involved in the match. Since player statistics change continuously it is an important quality control to know what the time interval is between when the match was played and when the player's statistics were fetched from the WOT API.

*Parsing the data stage 2: player's information*

In the player information parsing stage, player Id's were extracted from the replay data and queries about the statistics of these players were sent to the WOT API. The player information retrieved was then processed for relevant statistics (e.g. how good the player is) and persisted in the player information table in the database. Since the data collection process was carried out over several months (and will probably continue to be carried out for years in the future) the player statistics were connected to the replay they were associated with. By collecting statistics for every replay, players that were found in two different replays six months apart the system will have captured snapshot of that player's statistics for both points in time. Capturing multiple snapshots of a player's development also allows for more advanced avenues of analysis in the future i.e. is it possible that a player who is improving rapidly over time feels less frustrated than a player who has stagnated at the same playing level over the same period of time and could this lead to less negative chat behaviour?

*Parsing the data stage 2b: vehicle information*

Although the data was not used in this paper, the vehicle used by the player in the particular replay was also parsed and information about the vehicle was fetched from the WOT API and persisted in the database.

*Parsing the data stage 3: chat and other packet information*

The replay JSON data contains a large number of packets which update the client on all events that happen within the game. An event could be that the player's vehicle has been hit and damaged, that an enemy vehicle has been spotted or that a team mate has sent a chat message to the team. Private messages cannot be sent to an individual within a game so even messages sent to a particular player have to be broadcast to the whole team. The packet structure is shown in below.

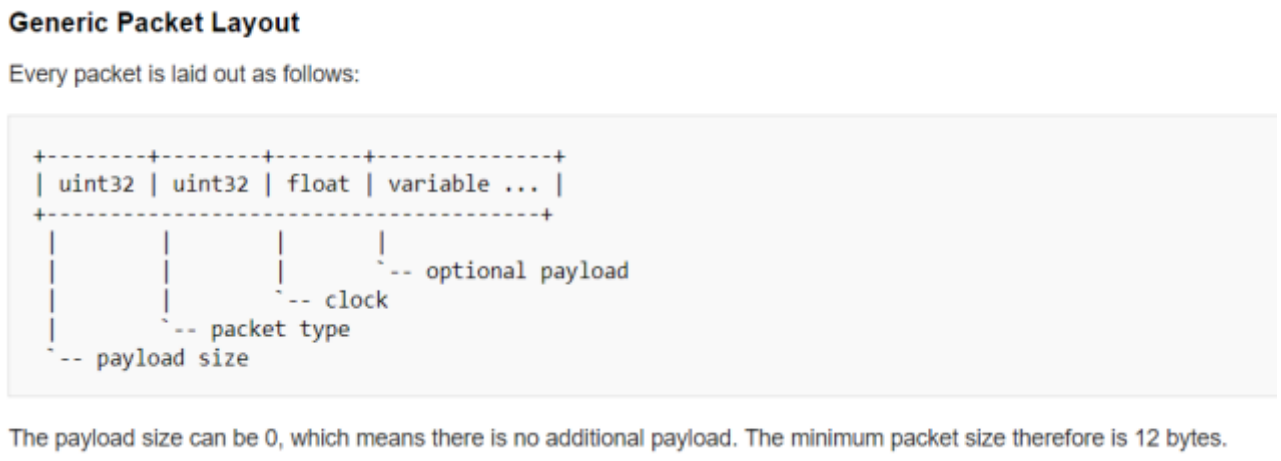

**Generic Packet Layout**

Every packet is laid out as follows:

```
+--------+--------+-------+-------------+
| uint32 | uint32 | float | variable ... |
+--------+--------+-------+-------------+
 |        |        |       |
 |        |        |       `-- optional payload
 |        |        `-- clock
 |        `-- packet type
 `-- payload size
```

The payload size can be 0, which means there is no additional payload. The minimum packet size therefore is 12 bytes.

**Figure 1 WOT replay packet structure (vbaddict, 2017)**

There are many different types of packets and not all the packet types have been identified by the developer community, however some of the known packet types are shown next:

| Packet Type | Description |
|---|---|
| 0x00 | Battle level setup |
| 0x01 | |
| 0x02 | |
| 0x03 | Tank appeared (spotting) |
| 0x04 | |
| 0x05 | Tank appeared (spotting) |
| 0x07 | Various tank related updates (health, tracks repaired etc.) |
| 0x08 | Various game state updates (player damages another player, player destroyed) |
| 0x0a | Vehicle position and rotation updates |
| 0x0b | |
| 0x0e | |
| 0x11 | |
| 0x12 | Game status |
| 0x13 | |
| 0x14 | The first packet you will see, and only see once |
| 0x16 | Recticle informations (position and ?) |
| 0x17 | |
| 0x18 | Possibly player camera angle (x) |
| 0x19 | Possibly player camera angle (y) |
| 0x1B | Possibly player camera angle (z) |
| 0x1C | |
| 0x1E | Main player id |
| 0x20 | |
| 0x21 | Minimap click |
| 0x22 | Main player specific data |
| 0x23 | Chat message |
| 0x25 | |
| 0x26 | Main player specific data |
| 0xffffffff | The last packet you will see, indicates the end of stream |

**Table 2 WOT replay packet types**

Two packet types are of particular interest in this data collection process: packets of types 35 (chat messages) and 8 (player updates). The type 35 messages contain in-game chat messages, while the type 8 messages contain (where the subtype is 44) the updates that designates when a player has died. An example of type 35 and type 8 packet data is shown below.

**Type 8 packet**

{"clock":112,"destroyed_by":17825077,"destruction_type":"shell","player_id":17686213,"sub_type":44,"target":17825104,"type":8},

Since this packet has a "destroyed by" field we can record the time of death for each player. In this example player 17686213 was killed at time 112 within the game.

**Type 35 packet**

{"clock":161,"message":"<font color='#80D63A'>Seco42[PCN] (T-150) :</font><font color='#80D63A'>M37 hit me</font>","type":35},

From this packet payload it can be determined that player "Seco42" sent the message "M37 hit me" at time 161 within the game.

At this stage in the replay parsing all packets are parsed, the time of death for each player are updated in the WotPlayer table and the chat messages are persisted to the ReplayMessages table.

### D. Data collection results

There were linits on how fast the system can collect data. This limit was imposed by the level of access given to the WOT API. Large scale queries to these services need to be approved with a commercial contract so in this data collection exercise the data gathering system processed just 1000 files on every data collection run. After running the system for a few months, the system has processed approximately 15,000 replays with approximately 44,000 associated player statistic snapshots and a total of approximately 126,000 messages.

| WOT OBJECT | TOTAL PROCESSED |
|---|---|
| **REPLAYS** | 14930 |
| **PLAYER STATISTIC SNAPSHOTS** | 443150 |
| **REPLAY MESSAGES** | 126091 |

**Table 3 Data collection results**

## IV. ANALYSIS

### A. Message classification scheme and score

Each message was assigned eight descriptive attributes that are potentially useful in terms of cyberbullying analysis.

- **IsAbusive**. If the content of the message could be considered having a negative emotional impact on the recipient or recipients. In this study, the variable is true when the message is considered to be a cyberbullying message.
- **IsPositive**. Is true if the emotional content of the message is generally positive, e.g. "well played" or "thanks for the help"
- **IsNegative**. This variable is probably unnecessary since it is the inverse of the last variable. When the IsPositive variable is true the the IsNegative variable must be false and vice versa.
- **HasBadLanguage**. If the message contains strong swear words. A list of the swear words considered strong are given in Appendix.
- **IsRacist**. When the message contains negative sentiments directed against a particular group on the basis of national identity, religious affiliation or sexual orientation.
- **NoobRelated**. When the message contains negative comments that suggest the player is playing so badly that they must be a very new and inexperienced player. A typical expression of this is calling another player a "noob" which is short for newbie or new player and has a negative connotation within a gaming context (Blackburn and Kwak, 2014).
- **SpecificTarget**. When the target of message is a single player, e.g. "Tiger tank you suck" or a small group of players "Our heavy tanks are useless", as opposed to a generally directed message "this team is useless". In this study messages directed against a specific target are considered to have a higher intensity of cyberbullying than those that are general.
- **FilteredText**. World of Tanks allows for the player to set filtering on for in game messages. Filtered text appears as stars, e.g. "you useless ****". Although the stars can be considered to be swear words it might be worth differentiating between the two.

*Cyberbullying score (CS).*

A cyberbullying score was calculated for each message based on the following points score

**For positive messages**
IsPositive -1pt
SpecificTarget -3pt

**For negative messages**
IsNegative +1pt
NoobRelated +1pt
HasBadLanguage or FilteredText +2pt
IsRacist +2pt
SpecificTarget +3pt

Every message will have as a score between -4 and +9 points. Continued bullying within the same match is considered as increasing the intensity so that if every negative message from the same author will add a further +1 points. The first negative message will then have +0 added to its score, the second +1, the third +2 etc. As an example, if a player typed 3 messages in chat in this order then we generate Table 4.

| Example message | Cyberbullying score (CS) | Explanation |
|---|---|---|
| Tiger tank you are useless | 4 | Negative(1), specific(3) |
| Go play with Barbie dolls u noob | 6 | Negative(1), specific(3), noob(1), 2nd message(1) |
| U useless **** tiger | 8 | Negative(1), specific(3), filtered(2), 3rd message(2) |

**Table 4 Scoring examples**

### B. Method

*Message classification (automatic)*

Having collected the messages, an initial naive automatic classification of was carried out using SQL queries e.g. NoobRelated is set to true if the message text contains the text "Noob". The purpose if this rough classification was mainly to facilitate manual classification and to facilitate some quick descriptive analysis of the data collected. Some examples of the SQL script used is given in the Appendix but it is also useful as a comparison against the potentially more sophisticated sentiment analysis techniques. Figure 2 shows the results of the automatic classification.

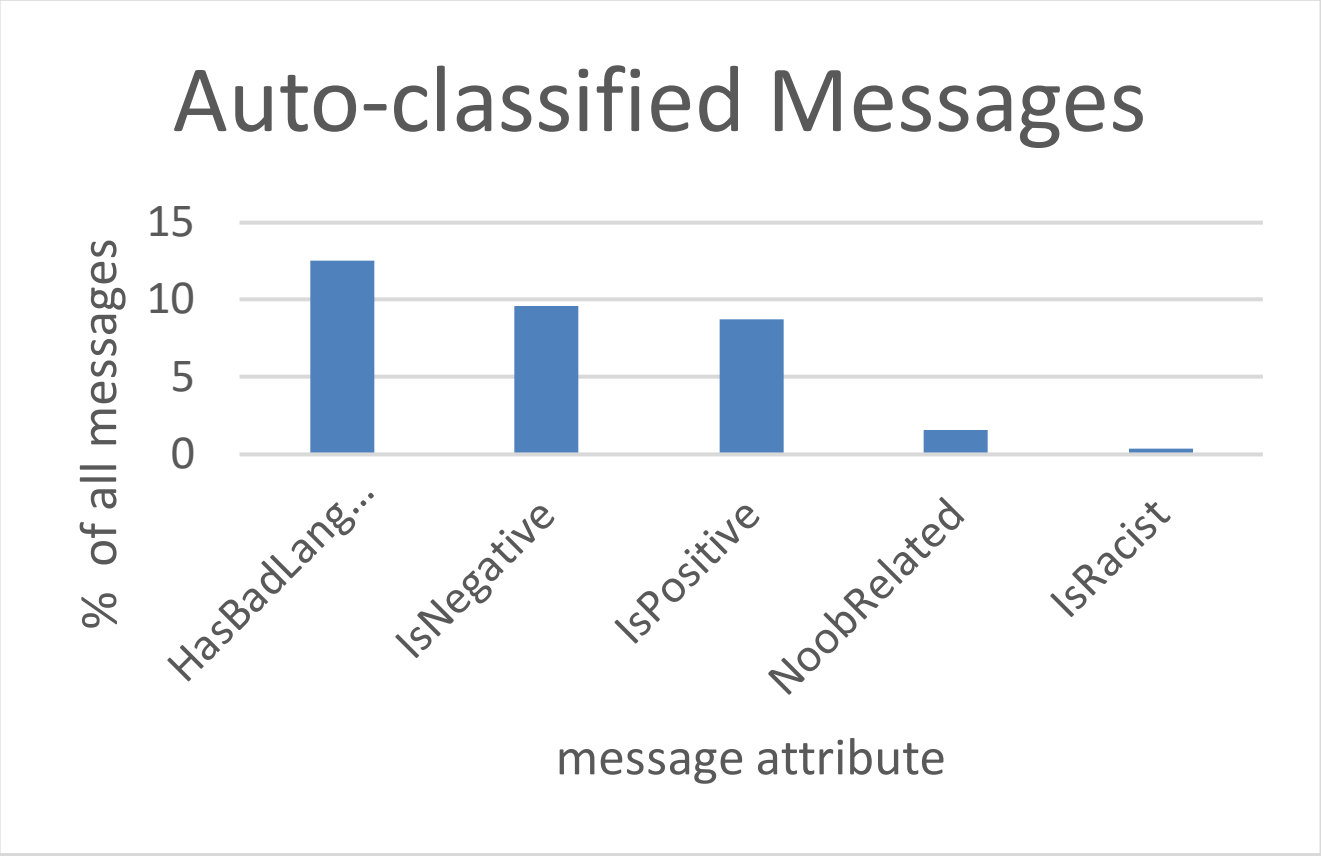

**Figure 2 Automatic classification result**

*Message classification (manual)*

A windows forms client was also built which allowed for the detailed manual classification of messages. The main GUI of the client is shown in Figure 3. The purpose of this client is to allow rapid classification of messages.

The main window shows all the messages from a particular match in the order that they appeared in the game. It is useful to see all the messages since it is difficult to correctly classify a message without the context from the messages that had appeared previously. Under the main window the currently selected message is shown. Beneath the message are the possible message classifications. All variables have three possible states checked (true), unchecked (false) or unknown (null). As shown in Figure 8, some of the boxes are already checked; these are the preliminary classifications from the automatic classification. The clear button shown bottom centre can be used to quickly uncheck all the unknown classifications (Racist, Specific Target and Has Filtered text in this example). The buttons on the bottom left and right of the window allow navigation between the previous/next match chats and the individual messages within a chat. The user can filter out matches that have previously been classified using this client by checking the “Only unclassified chats” checkbox. On the right of the window are shown the unique id of the message and of the match (replay file id). These ids were mainly used in the development of the client to check the data in the database before and after classification. The save button persists the classification to the database.

The client was used to manually classify around 5000 chat messages. Theses classifications could then be used to examine the performance of the simple semantic analysis techniques.

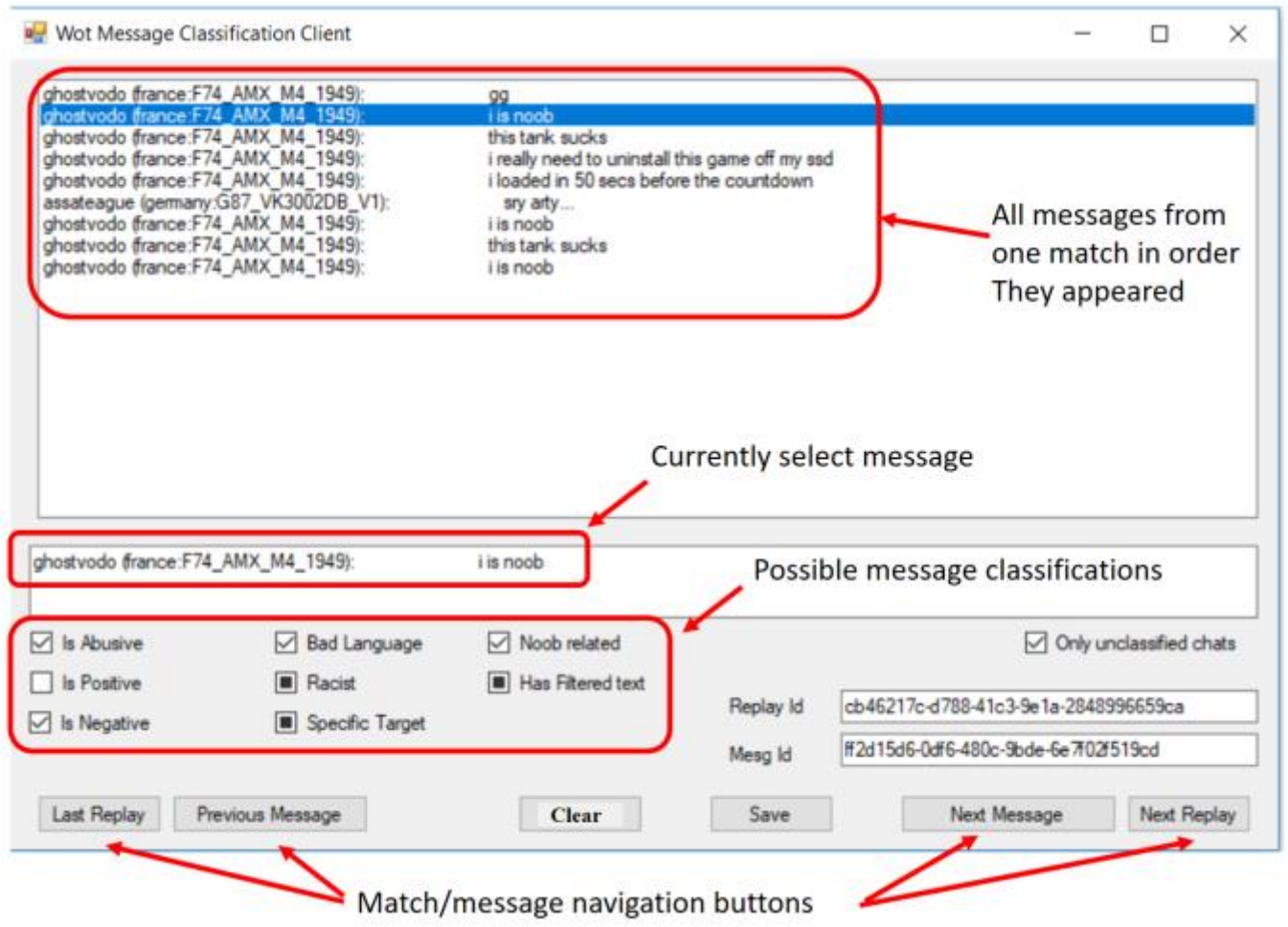

**Figure 3 Main GUI of the manual classification client**

*Relating player experience to cyberbullying behaviour*

We have a number of messages classified as “IsAbusive” through manual classification with the client previously. The player database also contains information on the player’s statistics fetched at the same time as the replay. The purpose of this is to do a preliminary study and see if Player XP can be a predictor in cyberbullying behavior (figures 8 and 9).

We are seeing a lot more messages, both abusive and non-abusive from players with less experience, this is presumably as a result of there being less players with higher levels of experience. In **Error! Reference source not found.**0 we can plot the statistic of the players that our data-collection system had fetched from Wargaming’s public API.

We can standardise the number of abusive messages per player against experience levels using this information. When this information is accounted for the result obtained is shown in **Error! Reference source not found.**.

It is difficult to draw conclusions about at the very high experience levels since the counts are so few, but at medium to low levels one can see that occurrence rates of cyberbullying messages are relatively constant or a slight trend upwards. It is a noteworthy feature that very low levels of cyberbullying messages come from players with very low levels of experience (less than 500K experience). In general, the diagram suggests that a player’s experience level may not be a great predictor of the probability of cyberbullying behaviour. It also suggests that players do not exhibit cyberbullying from the point they start playing the game. Cyberbullying seems to turn up after the players have been playing for a while. This could be an indication that cyberbullying within World of Tanks is a learned behaviour from other more experienced players.

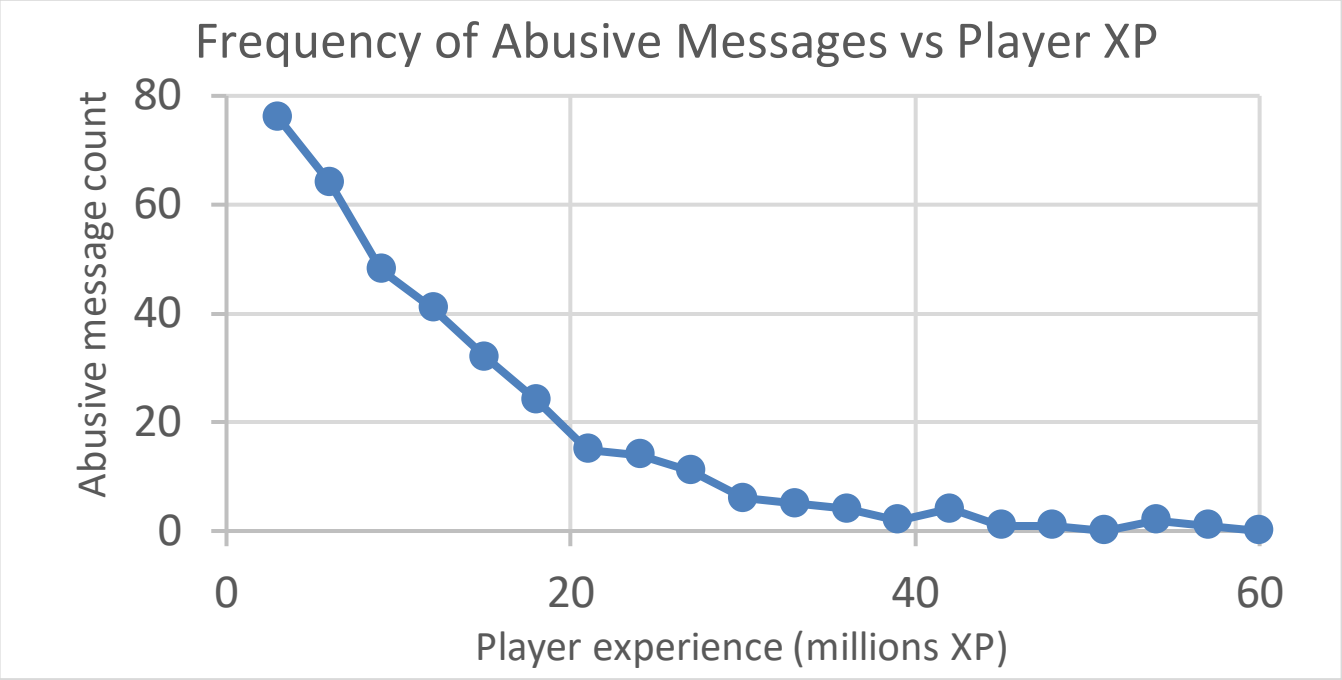

**Figure 4 Frequency of abusive message counts against player experience**

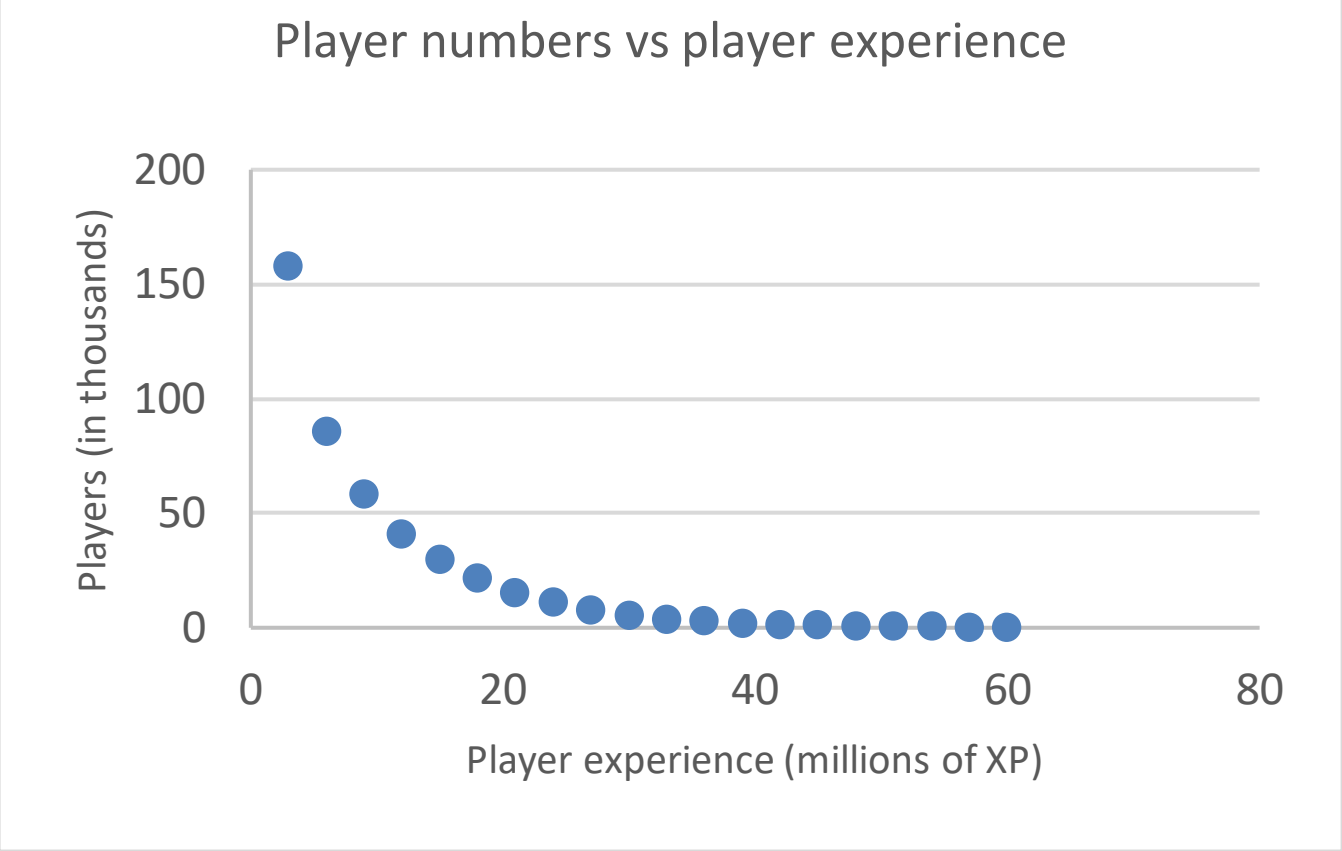

**Figure 5 Number of players vs player experience level**

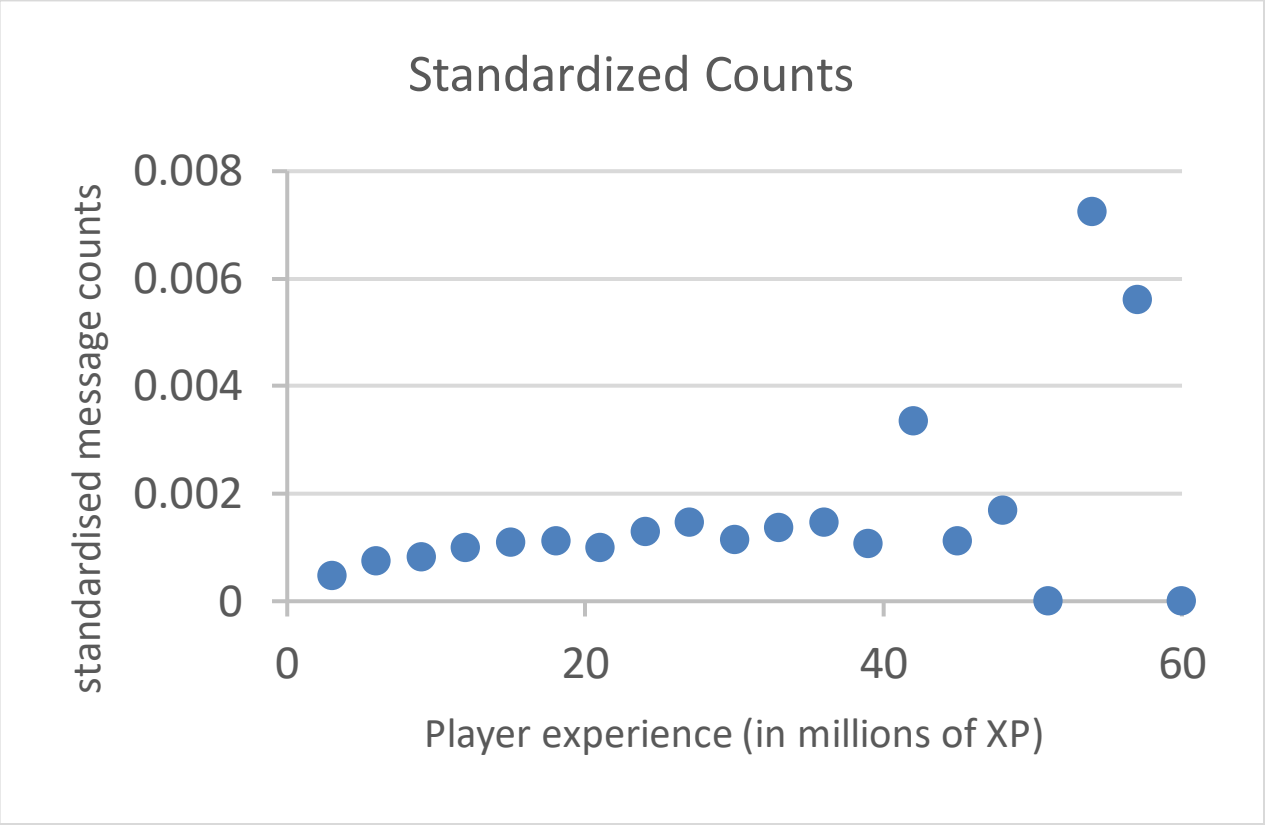

**Figure 6 Standardized abusive message counts vs player experience levels**

*Relating time of death in-game to cyberbullying behaviour*

One of the measurements that it was possible to extract from the chat data were the packets relating to the time of death of the players. Every chat message also has clock information. Using this clock information, it was possible to examine whether or not time of death had any relation to toxic messages. Looking at the messages that had been classified as IsAbusive and then comparing the chat message clock times to that player's time of death the it was found that 63% of all cyberbullying messages occurred after the player's death in the game. In it can be seen that most toxic messages occur a short period after the player's death.

This an interesting result. In a team game where players rely on their team mates but where the game mechanism does not necessarily reward team play then it is understandable that players might feel that their death was due to the action or lack of action by their team mates. At the point of death adrenaline and stress levels are probably higher, giving a higher probability of some sort of outburst.

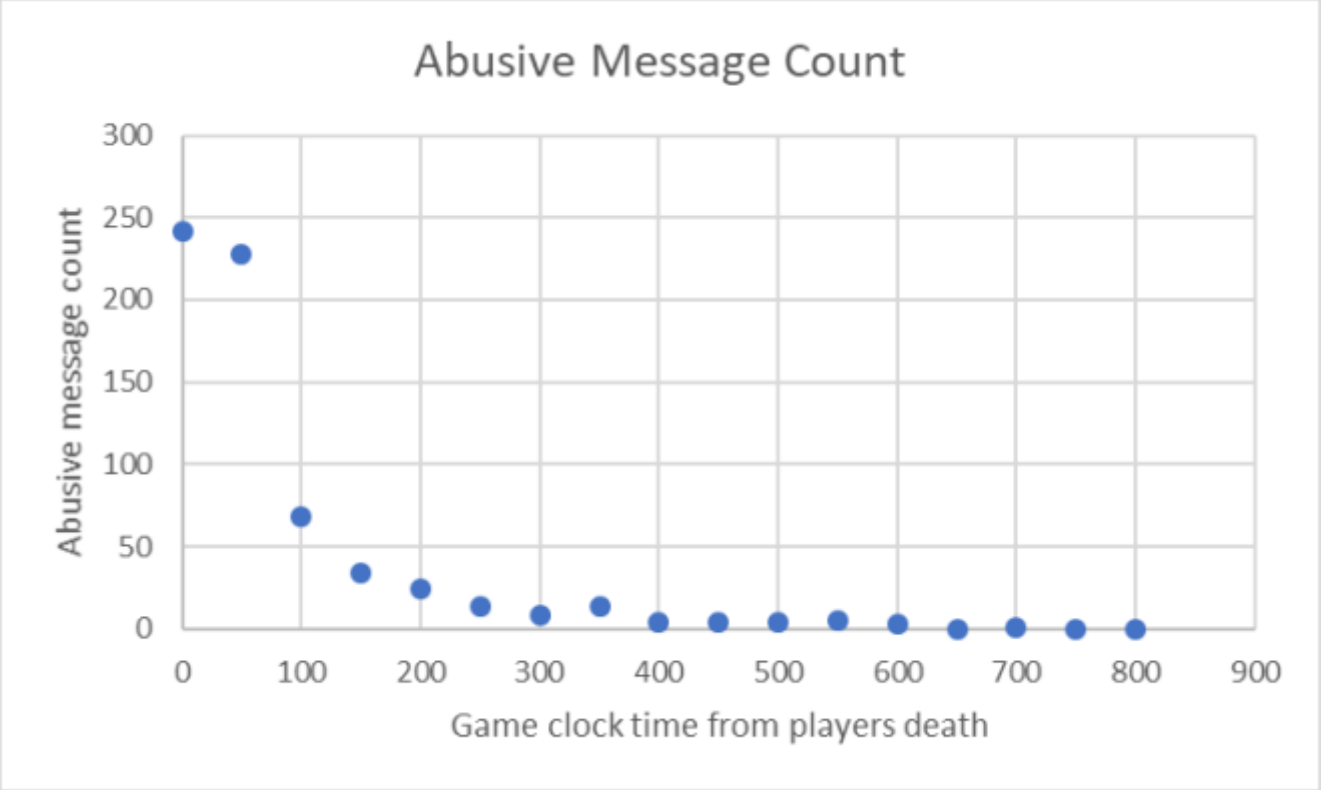

**Figure 7 Abusive message counts vs time after death**

*Sentiment analysis*

In this analysis two separate cloud-based sentiment analysis services were examined: Twinword Sentiment Analysis (Twinword, 2017) and Text Analytics from the Microsoft Azure Cognitive Services (Microsoft, 2017).

*Twinword Sentiment Analysis*

The Twinword sentiment analysis API offers only one method at the moment "analyse". Here is shown the results from two examples of text.

**Example 1 text to analyse:** I love ice cream!

**Analysis result**

```
{
  "type": "positive",
  "score": 0.917220858,
  "ratio": 1,
  "keywords": [
    {
      "word": "love",
      "score": 0.917220858
    }
  ],
  "version": "4.0.0",
  "author": "twinword inc.",
  "email": "feedback@twinword.com",
  "result_code": "200",
  "result_msg": "Success"
}
```

As can be seen the overall sentiment of the statement was judged to be positive with a score of 0.917 and a maximum ratio of 1 (ratio of positive scores to negative scores) since only 1

keyword was found (love) and this was determined to have a high overall positive score (0.917).

**Example 2 text to analyse:** I hate the whole team
**Analysis result**

```
{
  "type": "negative",
  "score": -0.383199971,
  "ratio": -0.71591411128435,
  "keywords": [
    {
      "word": "whole",
      "score": 0.152059727
    },
    {
      "word": "hate",
      "score": -0.918459669
    }
  ],
  "version": "4.0.0",
  "author": "twinword inc.",
  "email": "feedback@twinword.com",
  "result_code": "200",
  "result_msg": "Success"
}
```

In this result 2 keywords were found: “whole” (slight positive, 0.15) and “hate” (very negative, -0.918), with an overall negative result.

*Microsoft Text Analytics*

Microsoft’s Text Analytics is a little more sophisticated and offers four methods: detect language, detect key phrases and detect sentiment. Sentiment scores range from 0% which is negative to 100% which is positive.

**Example 1 text to analyse:** I love ice cream
**Analysis result**

```
{
  "languageDetection": {
    "documents": [
      {
        "id": "43c64c99-4cba-48e6-b79f-642bc413625d"
,
        "detectedLanguages": [
          {
            "name": "English",
            "iso6391Name": "en",
            "score": 1.0
          }
        ]
      }
    ],
    "errors": []
  },
  "keyPhrases": {
    "documents": [
      {
        "id": "43c64c99-4cba-48e6-b79f-642bc413625d"
,
        "keyPhrases": [
          "cream"
        ]
      }
    ],
    "errors": []
  },
  "sentiment": {
    "documents": [
      {
        "id": "43c64c99-4cba-48e6-b79f-642bc413625d"
,
        "score": 0.877212041746438
      }
    ],
    "errors": []
  }
```

| | |
|---|---|
| **LANGUAGES:** | English (confidence: 100%) |
| **KEY PHRASES:** | ice cream |
| **SENTIMENT:** | 88 % |

**Example 2 text to analyse:** I hate the whole team
**Analysis result**

```
{
  "languageDetection": {
    "documents": [
      {
        "id": "5268a45d-74b1-4dbb-99c6-a4b44f9da757"
,
        "detectedLanguages": [
          {
            "name": "English",
            "iso6391Name": "en",
            "score": 1.0
          }
        ]
      }
    ],
    "errors": []
  },
  "keyPhrases": {
    "documents": [
      {
        "id": "5268a45d-74b1-4dbb-99c6-a4b44f9da757"
,
        "keyPhrases": [
          "team"
        ]
      }
    ],
    "errors": []
  },
  "sentiment": {
    "documents": [
      {
        "id": "5268a45d-74b1-4dbb-99c6-a4b44f9da757"
,
        "score": 0.0478871469511812
      }
    ],
    "errors": []
  }
}
```

| | |
|---|---|
| **LANGUAGES:** | English (confidence: 100%) |
| **KEY PHRASES:** | team |
| **SENTIMENT:** | 5 % |

## V. Results

*Simple Naive Automatic Classification Performance*

**Figure 8 Automatic vs manual classification**

shows a quick comparison of the results of the manually classified (MAN) messages to the simple naïve automatically classified (SAC) messages.

There are a number of interesting points in this comparison that can be made on first inspection. Firstly: simple automatic classification (SAC) seems to give a very similar result to manual classification (MAN) when it comes to detecting noob related comments, racism or bad language, at least in the number of cases found within the chat. This might be because these results are defined by simple word detection. Secondly: MAN classification is clearly labelling a much higher percentage of cases as positive or negative than SAC classification. Finally: quite a high percentage of the messages are cyberbullying in nature (over 12%) and similarly many messages are directed against a specific target which suggests that cyberbullying is possibly quite prevalent within World of Tanks.

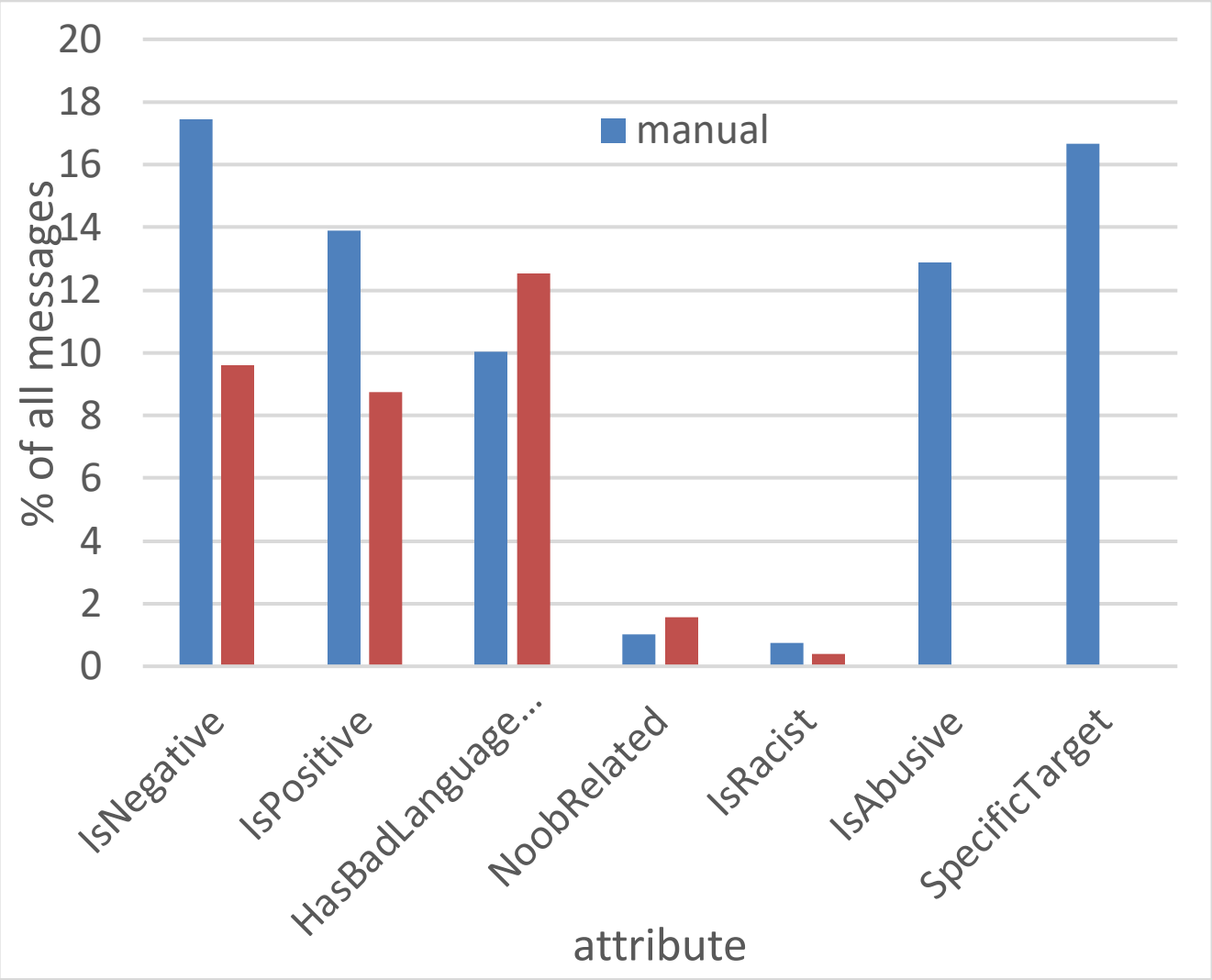

**Figure 8 Automatic vs manual classification**

Quick inspection is not however sufficient to really assess the quality of the SAC classification since, for example, even if the number of "noob" related messages found by both the MAN and SAC classification methods is similar there is no guarantee that both systems labelled the same cases. When measuring classification accuracy there are several single figure measures that can be used such as the diagnostic odds ratio (Glas et al., 2003) or the F-score (Visentini et al., 2016), but since this work was focused really on the proof of concept regarding the utility of the World of Tanks chat data, the diagnostic odds ratio (DOR) should prove sufficient to the task. By collecting the classified results, including false positives and false negatives we could calculate the DOR.

*IsRacist detection*

| | |
|---|---|
| MAN racist SAC racist | 18 |
| MAN non-racist SAC non-racist | 5021 |
| MAN racist SAC non racist (false negative) | 20 |
| MAN non racist SAC racist (false positive) | 2 |

**Figure 9 Classification results for IsRacist detection including false positives and negatives**

$$DOR = \frac{18 \times 5{,}021}{20 \times 2} = 2{,}259$$

The results look promising in regard to the low level of false positives, thus in an in-game experience very little useful communication would be blocked. However, it should be noted that SAC classification identified less than 45% of the manually identified cases.

By looking at some of the manually classified racist comments that were missed by the SAC: e.g. "your momy black pornstar", "hehe he likes bananas", "fresst scheiße ihr hirntoten judenschweine", it becomes possible to see why the SAC failed. The texts simply were not in the list provided. This suggests that the results for racist comment detection could be improved to a considerable degree by using a better indicator word list and possibly language detection/translation.

*HasBadLanguage detection*

| | |
|---|---|
| MAN bad SAC bad | 342 |
| MAN not bad SAC not bad | 4557 |
| MAN bad SAC not bad (false negative) | 150 |
| MAN not bad SAC bad (false positive) | 11 |

**Figure 10 Classification results for HasBadLanguage detection including false positives and negatives**

$$DOR = \frac{342 \times 4{,}557}{150 \times 11} = 946$$

With bad language detection again the Sac was quite successful. Examples of false negatives include: "fakking light tanks..", "and now you can kiss my back side idiot", "fffsss". Again the indications are that these results could be much improved by refining the word detection criteria.

*IsNegative detection*

| | |
|---|---|
| MAN neg SAC neg | 389 |
| MAN pos SAC pos | 4132 |
| MAN neg SAC pos (false negative) | 494 |
| MAN pos SAC neg (false positive) | 2 |

**Figure 11 Classification results for IsNegative detection including false positives and negatives**

$$DOR = \frac{389 \times 4{,}132}{494 \times 48} = 1{,}627$$

The result for classifying a message as having negative content are similar to previous categories. This is perhaps surprising since a negative sentiment is a much more diffuse, much harder to define concept than racist comments or bad language. Examples of false negatives include: "Fake ass game/", "teenager detected", "i think she is upset beacuse she has a son like u". Examples of false positives include: "stop spam ***" and "kidding right"

*NoobRelated detection*

| | |
|---|---|
| MAN noob SAC noob | 46 |
| MAN not noob SAC not noob | 5011 |
| MAN noob SAC not bad (false negative) | 5 |
| MAN not bad SAC bad (false positive) | 1 |

**Figure 12 Classification results for NoobRelated detection including false positives and negatives**

$$DOR = \frac{46 \times 5{,}011}{5 \times 1} = 46{,}101$$

One would expect the result to be high in this category since the feature depends on the existence of the word noob (or some variant) in the message text. False negatives include "to many tomato" and "nob med leo". False positives include "you nob". Although it is probably unnecessary the false negative could be improved by including the term "tomato" since this is a World of Tanks specific term for a new player. When player's statistics are showed they are usually colour-coded. Top players or unicums (Dictionary, 2017) statistics are highlighted in purple and while players with very poor statistics (often very new players or noobs) are highlighted in red, and are thus referred to as tomatoes (Room, 2017) .

*Repetition and specific targeting detection using SAC*

Using the SAC classification, it became possible to attempt to detect cyberbullying messages within the text. Two important factors involved in calculating the cyberbullying score (CS) have not been addressed so far by the SAC method: whether or not the comment in the chat is directed against a specific target and if the comment is part of a structure where the same person has made repeated comments within the same match. The first of these is probably beyond the ability of SAC and was not attempted within this work. Repeated cyberbullying comments may not be possible, but a weaker proxy measure is feasible. Using SQL it is possible to extract messages which have a negative nature, including racist, noob and bad language comments in grouped by comment source and replay/match. The extra scores can be applied to repetitive comments in this way. Doing this is is possible to update the repetition scores on each message classified by SAC. The script for carrying out this update is given in the Appendix.

*Calculating the cyberbullying score (CS) from the SAC and comparing to the results from manual classification.*

Using the previous measure, we can calculate a proxy cyberbullying score (PCS) similar to the CS score described previously. The score will be calculated according to the following formula:

```
PCS = (IsNegative * 1) + (NoobRelated *
1) + (HasBadLanguage|FilteredText * 2) +
(IsRacist * 2) + RepetiveMessageScore
```

Where the repetitive score was that calculated using the methods defined previously. The SQL script to update the score is given in Appendix. Figure 18 outlines the comparison of MAN cyberbullying score vs SAC proxy cyberbullying score.

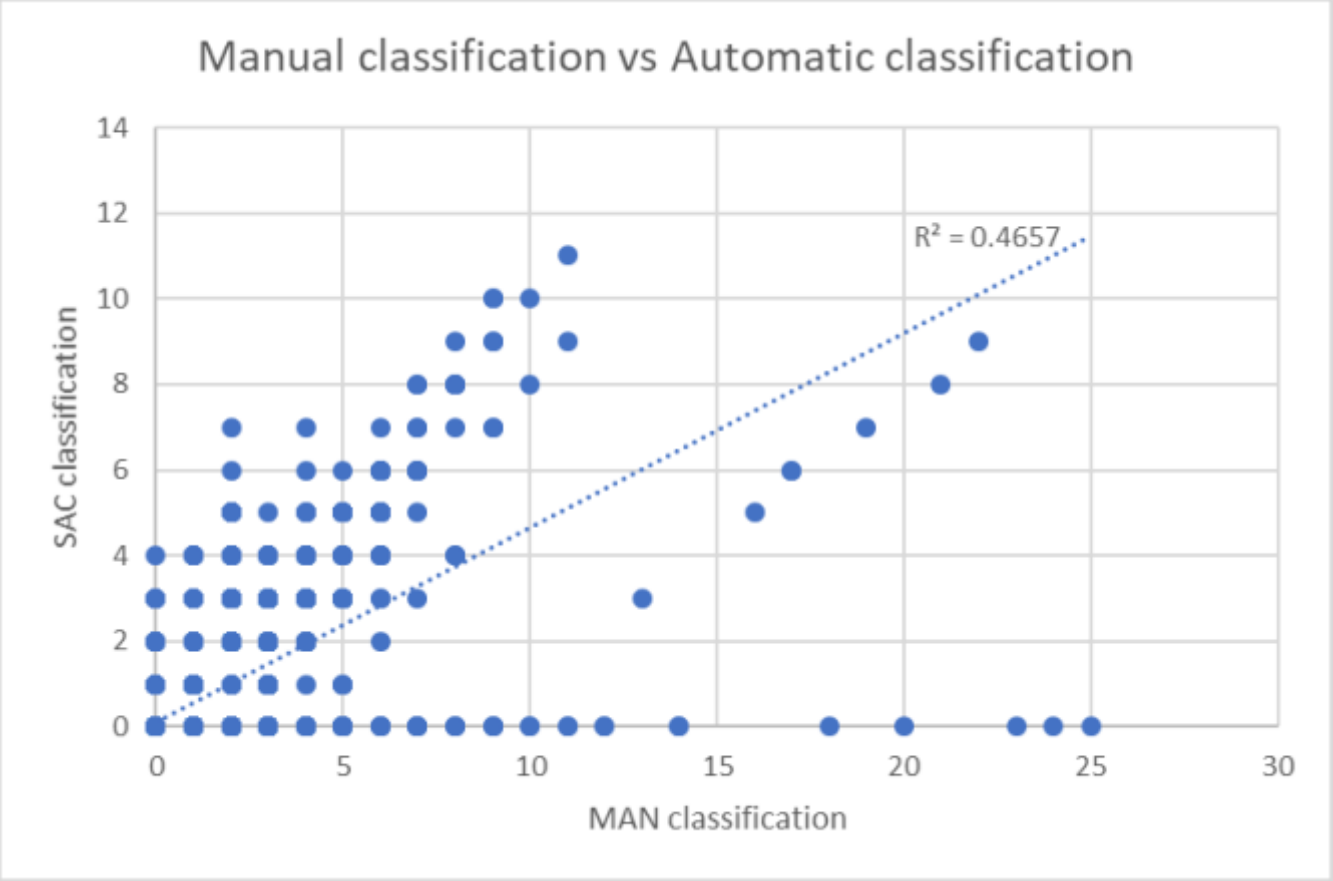

**Figure 13 MAN cyberbullying score vs SAC proxy cyberbullying score**

| | |
|---|---|
| MAN CS = 0 CAS PCS = 0 | 3987 |
| MAN CS > 0 CAS PCS > 0 | 654 |
| MAN CS > 0 CAS PCS = 0 (false negative) | 23 |
| MAN CS = 0 CAS PCS > 0 (false positive) | 399 |
| Total cases | 5063 |

**Table 5 Classification comparison of cyberbullying scores**

$$DOR = \frac{3987 \times 654}{23 \times 339} = 284$$

It is clear from the results that there is some utility in the simple naive classification method although it is also clear that it will not provide a full solution.

*Twinword sentiment analysis classification performance*

The manually classified messages were submitted to Twinword sentiment analysis service and the results were retrieved. Twinword classifies a sentiment as negative if the sentiment score is < -0.05 and positive if the sentiment score is > 0.05. The results are shown in Table 6.

| | |
|---|---|
| IsAbusive = 1 + negative sentiment | 309 |
| IsAbusive = 0 + positive sentiment | 1592 |
| IsAbusive = 1 + positive sentiment | 127 |
| IsAbusive = 0 + negative sentiment | 1279 |

**Table 6 Classification comparison of manual cyberbullying and Twinword sentiment analysis**

$$DOR = \frac{309 \times 1{,}592}{127 \times 1{,}279} = 3$$

The results from the Twinword sentiment analysis were surprisingly poor.

*Microsoft Azure sentiment analysis classification performance*

Here we examined the cases where MS text analytics found the chat statement to be positive or negative and compared that to whether or not the message had been classified as IsAbusive using the manual client.

| | |
|---|---|
| MAN not abusive MS positive | 204 |
| MAN abusive MS negative | 156 |
| MAN abusive MS positive (false negative) | 33 |
| MAN not abusive MS negative (false positive) | 145 |

**Figure 14 Classification comparison of manual cyberbullying and Microsoft sentiment analysis**

$$DOR = \frac{204 \times 156}{33 \times 145} = 6.6$$

The results from the Microsoft Azure sentiment analysis was also poor.

## VI. Discussion and conclusions

In this paper the current state of research into cyberbullying was examined. The serious nature of the problem was discussed considering the audience it affects and the fact that in many countries it is already or will probably be a specific cybercrime. A useful long term goal was identified: that of building a system that would be able to detect, counter or prevent cyberbullying when it occurs within a gaming environment. In order to achieve this goal several obstacles were identified. The first problem is the poor definition of cyberbullying and the overlap between this concept and similar concepts such as griefing.

Classification and scoring schemes that exist today were discussed and a scoring scheme suggested that is built around core cyberbullying features such as repetition and intensity. The scoring schema presented here is in no way meant to be a finished product, rather it was a useful measure to help assess the utility of the data gathered and the tools employed in this paper. It is certainly difficult to define one schema for cyberbullying that fits all circumstances since the different online environments have enormous differences in their circumstances e.g. an identity in Facebook is often much more tightly coupled to a real-life identity than an online gaming alter-ego. It may be useful therefore to define particular definition schemas particular to an environment and then attempt to find the common ground afterwards.

Another more serious problem that blocks progress in the building of an anti-cyberbullying system was identified: the lack of real data sets. In our review of current research, it was obvious that few researchers have access to large data sets to carry out their work. It was felt that this was one area where positive progress could be immediately made. World of Tanks was identified as a good initial target game for which we could start to collect data since there were 2 data sources that were publicly available, World of Tanks replay files from WotReplays.com and player statistics from the wargaming.net public API services.

A data collection system was designed that could scrape the link information from WotReplays.com, download, decode and transform the data to JSON data and then combine the information with player/vehicle information fetched from wargaming.net public services before persisting the results to a database.

The next stage in the paper is to examine the data we had collected to see if any useful insights could be obtained. To this end the data was classified using a simple set of SQL commandsand also using a custom designed classification windows client. The automatic classification results were compared with the manual results for the main attributes such as IsRacist, IsNegative. In general, the simple classification methods were more successful than expected. Although not perhaps a final solution it is likely that this low-level classification would be useful as a way of carrying out feature identification useful in combination with machine learning algorithms.

The chat data was also examined in combination with the player information from wargaming.net public services. Some interesting insights could be made from even this brief examination. Firstly, that cyberbullying did not seem to necessarily come from one particular group of players in terms of experience. More experienced players seem to engage in similar levels of cyberbullying to that of more junior players. It was notable however that this is not true for very new players suggesting that toxic behaviour is possibly a learned behaviour from other cyberbullies. This seems to at support some of the ideas in other research which suggests that cyberbullying spreads through communities in a similar manner to an epidemic (Fryling and Rivituso, 2013). Unfortunately, the size of the manually classified data was too limited to place much confidence in that conclusion, more data and further research is needed.

One interesting result the analysis of time-of-death and cyberbullying. Here there seems to be a very clear result that cyberbullying behaviour mostly occurs within a short time of death. As mentioned previously companies are aware of the problem and have attempted solutions such as filtering out swearing or blocking all in-game chat. These solutions simply don’t work since players find in-game chat useful and filtering does not prevent intimidation. The result shown in **Error! Reference source not found.** offers a very practical and possibly quite effective solution. If players are prevented from engaging in in-game chat for a short duration after their death it is quite feasible that in game toxic chat would be drastically reduced without any appreciable loss of useful communication. Implementation of such a feature would be a very simple matter for a gaming company and they have been known to act. World of Tanks, for example, removed the possibility of opposing teams being able to communicate with each other for the very reason that the chat became so toxic. This suggested solution has therefore the advantages of being feasible, practical and useful.

The final section using sentiment services offered the most interesting possibilities, but the results were however quite disappointing. Two possible reasons seem likely for this failure. The first and most likely is that the lack of experience in using these tools may have meant that the services were simply not used in a way that allowed them to operate to their full potential.

The second possibility is that the language structure and content of in-game chat is very specialised with often short code words used instead of full statements e.g. “gg” = good game, “wp” = well played or “typical lemming train” where a very stupid tactic is applied by a team with a resulting low chance of success. It is of course difficult for standard services to interpret this gaming language successfully. But it may be possible to adapt these services or create customised AI solutions that can do better. One area not examined by this research was the topics or key word functions available I Microsoft’s analysis services. It is possible that these functions can provide help in identifying the target of a particular chat conversation which would be useful with regard to the cyberbullying score. In our simple classification, we simply had no SQL that could identify if the comment (negative or positive) was directed against a specific target, which is a crucial factor where cyberbullying is concerned.

Although the work came nowhere near providing a system for detecting, countering or preventing cyberbullying it does open a number of interesting possibilities for future research and the data it will provide will certainly be of interest to researchers in the cyberbullying field.

Options for further research include extending the data gathering to non-English language users. This would require either some sort of translation method or alternatively updated classification methods that could handle non-English languages. Another avenue of further investigation is the use of sentiment analysis services. It seemed that the off the shelf text analysis services from Microsoft and Twinword had difficulty handling the in-game text. This might be because of the specialised nature of in-game chats which tend to be very brief and coded in gaming specific terms e.g. noob. It may be that a more specialised custom AI solution would provide much better results. The database should be a useful resource in building up a list of terms that gamers use.

## VIII. Appendix

All associated scripts, source code and swear words will be provided on a Web page:

http://asecuritysite.com/gamedata